\documentclass[twocolumn,superscriptaddress]{openjournal}

\usepackage{amsmath}
\usepackage{graphicx}% Include figure files
\usepackage{hyperref}% add hypertext capabilities
\usepackage{etoolbox}
\usepackage{orcidlink}
\usepackage{aas_macros}
\usepackage{xspace}
\usepackage{xcolor}

\newcommand{\vd}{\mathbf{d}}
\newcommand{\vm}{\mathbf{m}}
\newcommand{\vn}{\mathbf{n}} % Noise vector
\newcommand{\mA}{\mathrm{A}}
\newcommand{\mN}{\mathrm{N}} % Noise covariance
\newcommand{\mS}{\mathrm{S}} % Signal covariance
\newcommand{\mW}{\mathrm{W}} % Wiener filter matrix
\newcommand{\LN}{LuSEE-Night\xspace}
\newcommand{\vb}{\mathbf{b}}

\begin{document}

\preprint{arXiv:2508.16773}

\title{Linear map-making with LuSEE-Night}

\author{Hugo Camacho$^{1,*}$\orcidlink{0000-0001-5871-0951}}
\author{Kaja M. Rotermund$^{2}$\orcidlink{0000-0002-9181-9948}}
\author{An\v{z}e Slosar$^{1}$\orcidlink{0000-0002-8713-3695}}
\author{Stuart D. Bale$^{3,4}$\orcidlink{0000-0002-1989-3596}}
\author{David W. Barker$^{5}$\orcidlink{0009-0002-4218-4840}}
\author{Jack Burns$^{5}$\orcidlink{0000-0002-4468-2117}}
\author{Christian H. Bye$^{6,7}$\orcidlink{0000-0002-7971-3390}}
\author{Johnny Dorigo Jones$^{5}$\orcidlink{0000-0002-3292-9784}}
\author{Adam Fahs$^{3,2}$\orcidlink{0000-0002-7491-2753}}
\author{Keith Goetz$^{8}$\orcidlink{0000-0003-0420-3633}}
\author{Sven Herrmann$^{1}$}
\author{Joshua J. Hibbard$^{5}$\orcidlink{0000-0002-9377-5133}}
\author{Oliver Jeong$^{9,2}$\orcidlink{0000-0001-8101-468X}}
\author{Marc Klein-Wolt$^{10}$\orcidlink{0000-0001-7901-9545}}
\author{L\'{e}on V.E. Koopmans$^{11}$\orcidlink{0000-0003-1840-0312}}
\author{Joel Krajewski$^{4}$}
\author{Zack Li$^{12,2}$\orcidlink{0000-0002-0309-9750}}
\author{Corentin Louis$^{13}$\orcidlink{0000-0002-9552-8822}}
\author{Milan Maksimovi\'{c}$^{13}$\orcidlink{0000-0001-6172-5062}}
\author{Ryan McLean$^{4}$\orcidlink{0009-0006-6700-5692}}
\author{Raul A. Monsalve$^{4,14,15}$\orcidlink{0000-0002-3287-2327}}
\author{Paul O'Connor$^{1}$\orcidlink{0000-0002-8718-2235}}
\author{Aaron Parsons$^{6,7}$\orcidlink{0000-0002-5400-8097}}
\author{Michel Piat$^{9}$\orcidlink{0000-0003-0643-3088}}
\author{Marc Pulupa$^{4}$\orcidlink{0000-0002-1573-7457}}
\author{Rugved Pund$^{16,1}$\orcidlink{0009-0001-0044-4600}}
\author{David Rapetti$^{17,18,5}$\orcidlink{0000-0003-2196-6675}}
\author{Benjamin Saliwanchik$^{1}$\orcidlink{0000-0002-5089-7472}}
\author{Graham Speedie$^{16,1}$}
\author{Nikolai Stefanov$^{3,4}$}
\author{David Sundkvist$^{4}$\orcidlink{0000-0003-2794-7926}}
\author{Aritoki Suzuki$^{2}$\orcidlink{0000-0001-8101-468X}}
\author{Harish K. Vedantham$^{19,11}$\orcidlink{0000-0002-0872-181X}}
\author{Fatima Yousuf$^{3,4}$\orcidlink{0000-0002-3152-796X}}
\author{Philippe Zarka$^{13}$\orcidlink{0000-0003-1672-9878}}

\collaboration{LuSEE-Night Science Collaboration}

\email{$^*$hcamacho@bnl.gov}

\affiliation{$^1$Brookhaven National Laboratory, Upton, NY 11973, USA}
\affiliation{$^2$Physics Division, Lawrence Berkeley National Laboratory, Berkeley, CA 94720, USA}
\affiliation{$^3$Department of Physics, University of California, Berkeley, CA, 94720-7300, USA}
\affiliation{$^4$Space Sciences Laboratory, University of California, Berkeley, CA 94720-7450, USA}
\affiliation{$^5$Center for Astrophysics and Space Astronomy, Department of Astrophysical and Planetary Sciences, University of Colorado Boulder, CO 80309, USA}
\affiliation{$^6$Department of Astronomy, University of California, Berkeley, CA 94720, USA}
\affiliation{$^7$Radio Astronomy Laboratory, University of California, Berkeley, CA 94720, USA}
\affiliation{$^8$School of Physics and Astronomy, University of Minnesota, Minneapolis, MN 55455, USA}
\affiliation{$^9$Laboratoire Astroparticule et Cosmologie (APC), Université Paris-Cité, Paris, France}
\affiliation{$^{10}$Department of Astrophysics, Research Institute of Mathematics, Astrophysics and Particle Physics, Radboud University Nijmegen, Heijendaalseweg 135, 6525 AJ Nijmegen, The Netherlands}
\affiliation{$^{11}$Kapteyn Astronomical Institute, University of Groningen, P.O.Box 800, 9700 AV Groningen, The Netherlands}
\affiliation{$^{12}$Berkeley Center for Cosmological Physics, University of California, Berkeley, CA 94720, United States}
\affiliation{$^{13}$LIRA, Observatoire de Paris, Université PSL, Sorbonne Université, Université Paris Cité, CY Cergy Paris Université, CNRS, 92190 Meudon, France}
\affiliation{$^{14}$School of Earth and Space Exploration, Arizona State University, Tempe, AZ 85287, USA}
\affiliation{$^{15}$Departamento de Ingeniería Eléctrica, Universidad Católica de la Santísima Concepción, Alonso de Ribera 2850, Concepción, Chile}
\affiliation{$^{16}$Physics and Astronomy Department, Stony Brook University, Stony Brook, NY 11794, USA}
\affiliation{$^{17}$NASA Ames Research Center, Moffett Field, CA 94035, USA}
\affiliation{$^{18}$Research Institute for Advanced Computer Science, Universities Space Research Association, Washington, DC 20024, USA}
\affiliation{$^{19}$ASTRON, Netherlands Institute for Radio Astronomy, Oude Hoogeveensedĳk 4, Dwingeloo, 7991 PD, The Netherlands}

\date{\today}% It is always \today, today,

\begin{abstract}
\LN is a pathfinder radio telescope on the lunar far side employing four 3-m monopole antennas arranged as two horizontal cross pseudo-dipoles on a rotational stage and sensitive to the radio sky in the 1-50 MHz frequency band.
\LN measures the corresponding 16 correlation products as a function of frequency.
While each antenna combination measures radiation coming from a large area of the sky, their aggregate information as a function of phase in the lunar cycle and rotational stage position can be deconvolved into a low-resolution map of the sky.
We study this deconvolution using linear map-making based on the Wiener filter algorithm.
We illustrate how systematic effects can be effectively marginalised over as contributions to the noise covariance and demonstrate this technique on beam knowledge uncertainty and gain fluctuations.
With reasonable assumptions about instrument performance, we show that \LN should be able to map the sub-50 MHz sky at a $\sim$5-degree resolution.
\end{abstract}

\maketitle

\section{Introduction}

The lunar far side has been touted as the ultimate radio observatory for low-frequency ($<30$\,MHz) observations of the sky.
At these low frequencies, observations from Earth are limited by both the Earth's ionosphere, which refracts and reflects radiation from the sky, and emissions from anthropogenic and meteorological sources.
The lunar far side, permanently shielded from Earth's signals and, during the lunar night, also shielded from the Sun, has long been considered an ideal location for low-frequency radio receivers.
Indeed, there are several proposals for large futuristic telescopes, such as FarSide~\cite{2021arXiv210308623B}, ALO-DEX~\cite{2025arXiv250403418B}, LCRT~\cite{lcrt}, DSL~\cite{2021RSPTA.37990566C}, as well as several recent operating pathfinders, such as the receiver on Chang'e-4~\cite{2019EPSC...13..529S}, ROLSES~\cite{2021PSJ.....2...44B,2025arXiv250309842H} on the Intuitive Machines 1 mission, and the upcoming \LN on the Firefly's Blue Ghost 2 mission.

\LN is a pathfinder radio telescope slated for delivery to the lunar far side in September 2026. A preliminary instrument description can be found in~\citep{2023arXiv230110345B}, with a more complete instrument design paper in preparation.
Since the monopole antennas are electrically short at the lower end of the observed frequency range and have inherently low directivity, the instrument is more akin to a radio receiver than a traditional telescope.
Given the available 16 correlation products, coupled with lunar rotation and rotation of the antenna turntable, however, the dataset will contain a large number of distinct measurements.
Although each one of these measurements probes large patches of the sky, their number, and signal-to-noise will overconstrain the degrees of freedom available in the low-resolution map of the sky.
The purpose of this paper is to investigate this problem quantitatively and to determine if \LN's data can be used to recover a map of the radio sky.

To this end, we employ the Wiener filter, a well-established linear method for map reconstruction.
Wiener filtering provides the maximum a posteriori (MAP) map estimate assuming a Gaussian distribution for the input map pixels and noise.
It has been traditionally used in Cosmic Microwave Background (CMB) map-making~\cite{1997ApJ...480L..87T} and for general data compression in cosmology (e.g., \cite{1998ApJ...503..492S}).
In this work, we utilize Wiener filtering as a deconvolution process under a minimal set of assumptions.
Future work (e.g., \cite{Li2026}) will explore more aggressive non-linear map-making algorithms that may produce higher fidelity maps, albeit at the cost of stronger, and potentially incorrect, assumptions.

Conceptually, Wiener deconvolution is a linear process that optimally balances information from the data with prior knowledge.
On signal-dominated scales, it trusts the data to faithfully reproduce sky features.
Conversely, on noise-dominated scales, it regularizes the solution by pulling map modes (e.g., spherical harmonic coefficients or pixel values) towards their prior mean (typically zero, as in the Gaussian likelihood, zero is the most likely value in the absence of other information).
In the intermediate regime, it gracefully interpolates between these two extremes.
Consequently, the resulting map reconstructs modes that are sufficiently non-degenerate in the data space and possess adequate signal-to-noise.
While low-frequency radiometers are often strongly signal-dominated, a significant practical challenge is the precise knowledge of the antenna beam profiles.
In fact, as is typical in any deconvolution, the kernels need to be known precisely in order to correctly deconvolve the data.

In this paper, we limit our discussion to unpolarized sources and a deconvolution that operates one frequency at a time.
An extension to polarized sources and multiple frequencies, while more cumbersome, is not fundamentally different in its approach.

\section{Map-making with a short-antenna radio instrument}

\subsection{Observables}
\label{sec:observables}

Consider an array of $N$ independently amplified antennas, indexed $i=1, \ldots, N$.
These antennas produce time-varying signals that serve as input to a spectrometer.
After spectral channelization, the signal from antenna $i$ in a given frequency bin is described by a complex number $s_i$, encoding its amplitude and phase.
Since we treat each frequency bin independently in this paper, we will omit the frequency index for brevity.

For thermal radiation, all relevant information is contained in the two-point statistics of these complex signals.
We can form $N$ real-valued auto-correlation products, $V_{ii} = s_i s_i^*$, and $N(N-1)/2$ complex-valued cross-correlation products, $V_{ij} = s_i s_j^*$ (for $i \neq j$).
Each complex cross-correlation $V_{ij}$ provides two real degrees of freedom (its real and imaginary parts).
Therefore, the total number of independent real-valued observables is $N^2$.

The directional response of a given antenna is characterized by its complex electric field pattern, $(E_\theta(\hat{\mathbf{n}}), E_\phi(\hat{\mathbf{n}}))$, which describes its sensitivity to the $\theta$ and $\phi$ components of an incident electric field from direction $\hat{\mathbf{n}} \equiv (\sin\theta\cos\phi, \sin\theta\sin\phi, \cos\theta)^T$.
For any given direction $\hat{\mathbf{n}}$, these two complex components correspond to four real degrees of freedom, determining the antenna's sensitivity to the four Stokes parameters $(I, Q, U, V)$ of radiation arriving from that direction.

For a pair of antennas, with electric field responses $(E_\theta, E_\phi)$ for the first and $(E'_\theta, E'_\phi)$ for the second, the combined beam responses to the Stokes parameters are given by:
\begin{eqnarray}
        B_I &=& \frac{1}{2}\left(E_\theta E_\theta'^* + E_\phi E_\phi'^* \right) \label{eq:beam1}\\
        B_Q &=& \frac{1}{2}\left(E_\theta E_\theta'^* - E_\phi E_\phi'^* \right) \\
        B_U &=& \left(E_\theta E_\phi'^* + E_\phi E_\theta'^*\right)\\
        B_V &=& i (E_\theta E_\phi'^* - E_\phi E_\theta'^*)
\end{eqnarray}

For a radiation field described by Stokes parameters $(I,Q,U,V)$ from direction $\hat{\mathbf{n}}$, the contribution to the measured correlation product $V_{ij}$ is obtained by integrating the product of the appropriate beam response and the Stokes parameter over all solid angles.
Assuming unpolarized radiation ($Q=U=V=0$), the observed signal for antenna pair $(i,j)$ simplifies to an integral of the sky intensity $I(\hat{\mathbf{n}})$ weighted by the intensity beam:
\begin{equation}
        V_{ij} = \int B_{ij} (\hat{\mathbf{n}}) I(\hat{\mathbf{n}}) d\Omega, \label{eq:beam2}
\end{equation}
where $B_{ij}$ denotes the intensity beam pattern (equivalent to $B_I$ in Eq.~\ref{eq:beam1}) for the antenna pair $(i,j)$;
for brevity, we have dropped the $I$ subscript.

\subsection{Electrically uncoupled identical antennas}

The formalism presented up to this point is general and does not assume specific properties for the antennas or their coupling, provided all beam responses are defined with respect to a common phase center.
A common simplification arises in the ideal case of two identical, electrically uncoupled antennas separated by a baseline vector $\vb_{ij}$.
In this scenario, if the electric field response of the first antenna is $E(\hat{\mathbf{n}})$, the response of the second antenna (displaced by $\vb_{ij}$) is $E'(\hat{\mathbf{n}}) = E(\hat{\mathbf{n}}) e^{i\mathbf{k} \cdot \vb_{ij}}$, where $\mathbf{k} = (2\pi/\lambda)\hat{\mathbf{n}}$ is the wavevector in direction $\hat{\mathbf{n}}$.
This consideration leads to the standard interferometric visibility expressions:
\begin{eqnarray}
        V_{ii} = V_{jj} = \int B(\hat{\mathbf{n}}) I(\hat{\mathbf{n}}) d\Omega, \\
        V_{ij} = \int B(\hat{\mathbf{n}}) I(\hat{\mathbf{n}}) e^{i\mathbf{k} \cdot \vb_{ij}} d\Omega,
        \label{eq:interferometer}
\end{eqnarray}
where $B(\hat{\mathbf{n}})$ is the intensity beam pattern of a single, isolated antenna.
Here, $B(\hat{\mathbf{n}})$ is assumed to be a real-valued function, identical for all antennas, representing the primary beam.

However, for this simplified interferometric limit (Eq.~\ref{eq:interferometer}) to accurately describe the observations, two conditions must be met:
\begin{itemize}
        \item The intrinsic beam patterns of the two antennas must be identical.
        \item The antennas must be electrically uncoupled; i.e., exciting one antenna should not induce significant currents or fields in the other.
\end{itemize}

In standard interferometers, the first condition is met by building telescopes with identical elements and the second condition is met by sufficiently separating the elements so as to electrically uncouple them.
It is important to note that these idealized conditions are generally \emph{not} met for an instrument like \LN.
The antennas are not identical (nor oriented identically) and are spaced closely enough to be electrically coupled.
Therefore, instead of using the simplified interferometric model, it is more accurate to treat each of the 16 real-valued correlation products as an independent observable, each with its own unique beam response function:
\begin{equation}
        V_{k} = \int B_k(\hat{\mathbf{n}}) I(\hat{\mathbf{n}}) d\Omega, \label{eq:beam3}
\end{equation}
with $k=1\ldots 16$ and $B_k(\mathbf{n})$ strictly real functions of sky position.
This is the formalism discussed in Section~\ref{sec:observables}, which accounts for distinct beam patterns $B_{k}$ for each correlation pair.
Nevertheless, we will still use the word ``visibility'' for $V_k$, which is admittedly mauling of its etymological origin.
\begin{figure}
        \centering
        \includegraphics[width=\linewidth]{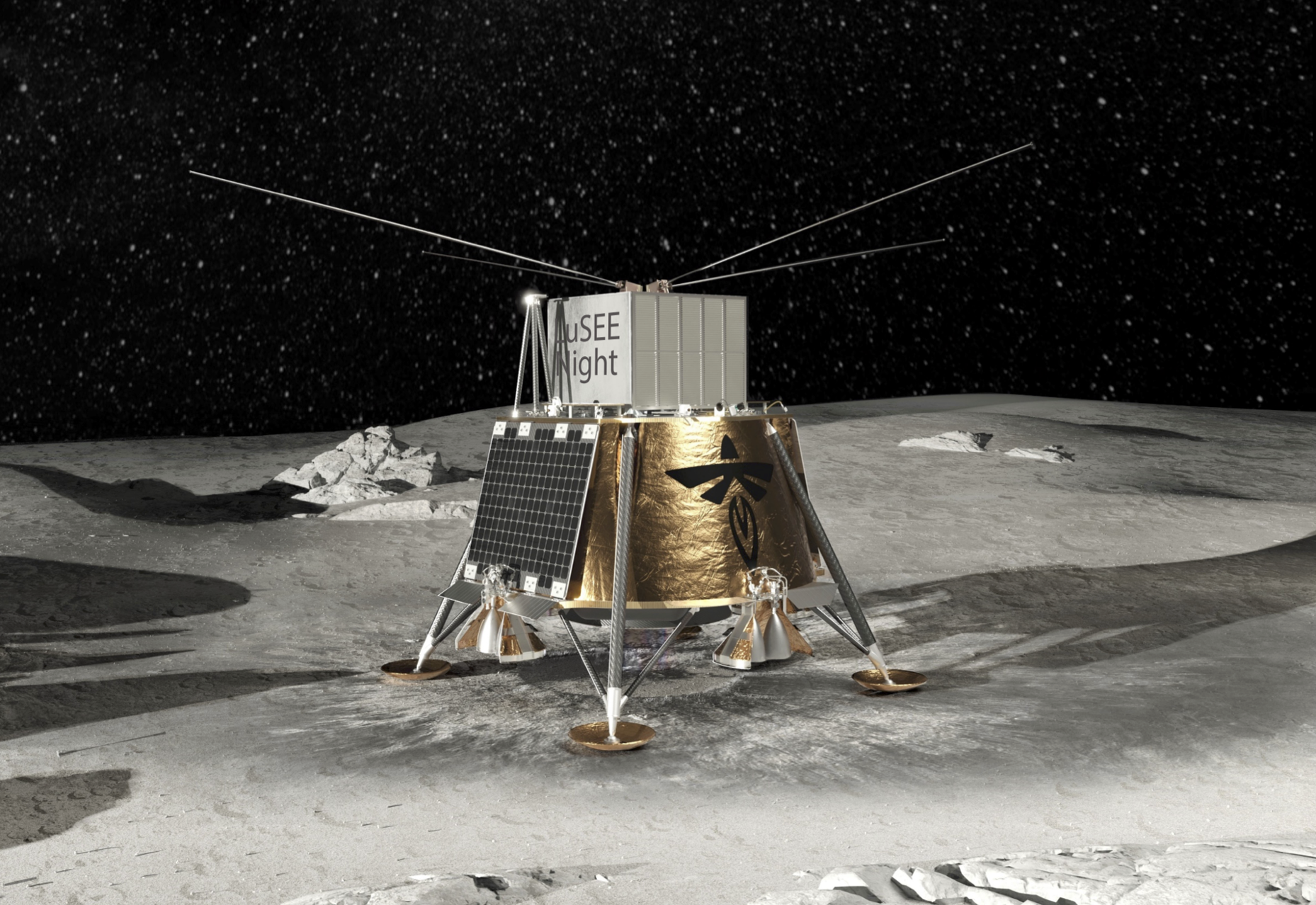}
        \caption{Rendering of the LuSEE-Night instrument on top of the Blue Ghost Mission 2 Lander.
Four monopole stacer antennas can be observed on the top.
Rendering courtesy of Firefly Aerospace.}
        \label{fig:bg2}
\end{figure}

\subsection{LuSEE-Night}

\LN is a radio pathfinder instrument manifested to land on the lunar far side in September 2026 as part of the Commercial Lunar Payload Services (CLPS) CS-3 mission.
The instrument is depicted in Figure~\ref{fig:bg2}. It consists of four monopole antennas, each approximately 3\,m long, forming a cross pseudo-dipole pair.
The entire antenna subsystem is situated on a rotational turntable, which can be actuated during the lunar day (which lasts approximately 14 Earth days) to change the orientation of the antennas with respect to the sky.
\LN is designed to survive multiple lunar nights and observe the sky as it transits overhead, effectively acting as a transit radio telescope.

Each monopole is independently amplified. The core of the instrument's digital system is a 4-channel spectrometer and correlator.
This system channelizes the input signal from each antenna into 2048 spectral bins, covering the frequency range of 0--51.2\,MHz with an inter-bin spacing of 25\,kHz.
For these four input signals, the correlator calculates all four auto-correlation products ($V_{ii}$) and all six unique complex cross-correlation products ($V_{ij}$, $i \neq j$).
As discussed above, despite a superficial similarity, it is more accurate to consider \LN not as a traditional interferometer but as an instrument measuring a set of beam-weighted sky integrals.
At any given time, the instrument measures 4 real-valued auto-correlations and 6 complex-valued cross-correlations.
This provides a total of $4 + 2 \times 6 = 16$ independent real-valued observables as described by Equation~\ref{eq:beam3}.
In the instrument's reference frame, the antenna beam patterns are fixed (for a given turntable angle), and the sky rotates due to lunar motion.
Conversely, in the sky's reference frame, the celestial signal is static (ignoring transient sources), and the instrument's beams scan across the sky due to both lunar rotation and any commanded turntable rotation.

Because the antennas are small compared to the observed wavelengths, each individual antenna beam pattern covers a large solid angle on the sky.
Nevertheless, the combination of 16 independent correlation measurements at each time step, coupled with the changing sky illumination due to lunar rotation and potential turntable adjustments, provides a diverse set of constraints.
This rich dataset makes it possible to deconvolve the sky signal and reconstruct a map using appropriate map-making methods, as the system can become over-constrained.

\subsection{Simulating the data vector}

The simulations in this work are based on the \texttt{luseepy}\footnote{\url{https://github.com/lusee-night/luseepy}} Python package, which provides the necessary tools to generate realistic data vectors for \LN.

For the input sky model, we use the Ultra-Low frequency Sky Model (ULSA)~\cite{2021ApJ...914..128C}, which we regenerate at frequencies from 1 to 50\,MHz with a 1\,MHz spacing.
We assume the sky is completely unpolarized. This assumption is justified because significant Faraday rotation is expected to erase any polarization response for the typical bandpass in spectral bins accessible to \LN.
We therefore simulate only the Stokes $I$ parameter.

For the prediction of the beams we use the High-Frequency Structure Simulator HFSS\footnote{\url{https://www.ansys.com/products/electronics/ansys-hfss}} simulation of a single \LN monopole.
The primary input for the beam model is the simulated complex electric field pattern $(E_\theta(\hat{\mathbf{n}}), E_\phi(\hat{\mathbf{n}}))$ of an individual \LN monopole antenna, obtained from electromagnetic simulations using HFSS.
Although \LN is not strictly four-fold symmetric due to the presence of asymmetries in the lander design, the additional lander payloads and \LN radiator, it is close to being four-fold symmetric.
For the purpose of this investigation, we take an electromagnetic simulation of a single active \LN monopole, with the other three monopoles included as passive elements, and rotate it repeatedly by 90$^\circ$ in order to generate $E$-field responses for all monopoles.
This allows us to construct the intensity beam patterns $B_{ij}(\hat{\mathbf{n}})$, Eq. (\ref{eq:beam1}), for any pair of antennas including auto-correlations.

\begin{figure*}
        \centering
        \includegraphics[width=\linewidth]{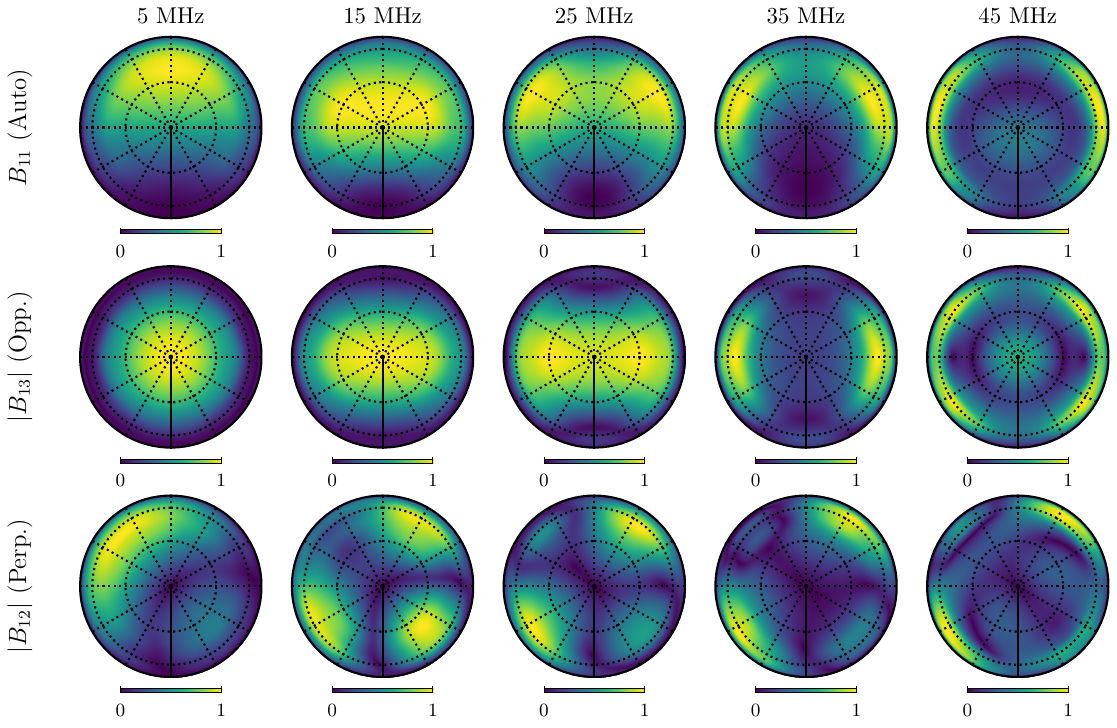}
        \caption{Simulated beam patterns derived from HFSS simulations at 5, 15, 25, 35, and 45 MHz (columns left to right).
Top row: Auto-correlation power pattern $B_{11}(\theta, \phi)$ for \LN antenna 1 (nominally North).
Middle row: Magnitude of the cross-correlation beam pattern $|B_{13}(\theta, \phi)|$ for the opposite antenna pair (1-3).
Bottom row: Magnitude of the cross-correlation beam pattern $|B_{12}(\theta, \phi)|$ for the perpendicular antenna pair (1-2).
Projection: Orthographic, zenith-centered ($\theta=0^\circ$), showing the hemisphere above the lunar surface.
                Orientation: North (antenna 1 axis projection) is up.
Scale: All beam patterns are normalized to their peak intensity at each frequency to show relative structure (0 to 1, arbitrary units).
Note the broad spatial response and its significant broadening towards lower frequencies.
Beyond 25\,MHz the antenna resonance causes the beam to bifurcate.}
        \label{fig:beam_power}
\end{figure*}

Figure~\ref{fig:beam_power} illustrates the resulting auto-correlation power beam pattern $B_{11}(\hat{\mathbf{n}})$ for antenna 1 (nominally oriented North), as well as the magnitude of the cross-correlation beam patterns for the opposite ($|B_{13}|$) and perpendicular ($|B_{12}|$) antenna pairs, at several representative frequencies.
These HFSS-derived patterns exhibit the expected broad spatial response characteristic of electrically short antennas, particularly at lower frequencies where the beam significantly broadens.
While each individual measurement integrates signal over a large sky patch, the instrument's sky scanning strategy (lunar rotation combined with potential turntable adjustments) systematically varies the beam's orientation relative to the celestial sphere.
This modulation provides the distinct, overlapping views necessary for the map-making algorithm to deconvolve the spatial structure from the temporally sequenced measurements.

A careful reader might ask, given the four-fold symmetry of the lander, whether the {11} (North-North) and {33} (South-South) correlations do not measure the same quantity.
In fact, as Figure \ref{fig:beam_power} shows, the beam for {11} combination is skewed North and is therefore not degenerate with {33} combination.
Furthermore, one might ask if {12} (North-East) correlation is not sensitive to U and V polarization, which are assumed zero in the current model?
The answer is that a pair of cross-dipoles is only a polarimeter for source at zenith.
Sources lower on the horizon can produce those correlations even when only $I$ polarization is present.
In short, all 16 correlation products are non-zero and independent measurements of the sky signal, although some are naturally larger than the others.

The \texttt{luseepy} package combines the sky model and the beam patterns to compute the simulated data vector $\vd$.
This forward model implicitly includes system gains, which for the fiducial simulation are assumed to be perfectly known and stable.
The impact of gain fluctuations is treated as a systematic effect in Section~\ref{sec:gain_fluctuations}.
The package utilizes the \texttt{lunarsky}\footnote{\url{https://github.com/aelanman/lunarsky}} package to accurately determine the orientation of the Moon, and thus the instrument, with respect to the celestial sphere at any given time.

At low radio frequencies, the system temperature is expected to be dominated by the Galactic synchrotron emission itself.
We modeled the noise variance for each visibility measurement using the radiometer equation appropriate for correlation products \cite{2016era..book.....C}:
\begin{equation}\label{eq:radiometer}
        \sigma_{ij}^2 = \frac{T_{ii} T_{jj} + T_{ij}^2}{2\Delta f \Delta t},
\end{equation}
where $T_{ij} \approx \langle V_{ij} \rangle$ represents the effective system temperature associated with the correlation product $V_{ij}$ (for auto-correlations $T_{ii}$, this is the system temperature of antenna $i$, dominated by the sky temperature; for cross-correlations, $T_{ij}$ is the correlated part of this temperature).
$\Delta f = 1$\,MHz is the assumed effective channel bandwidth for this simulation.
We assumed the noise is uncorrelated between different samples and frequency bins.

For the simulations in this paper, we assume \LN operates from the expected \LN landing site at the lunar far side at lunar longitude  182.258$^\circ$ and latitude -23.814$^\circ$.
Our fiducial simulation covers one full lunar sidereal rotation period (approximately 27.3 Earth days), sampling the 16 real-valued visibility components every $\Delta t = 7200$\,s (2 hours).
This corresponds to a continuous observation over a full lunar cycle.

While the instrument is designed to operate during both lunar day and night, the primary science operations are planned for the lunar night to minimize solar interference.
The choice of a full lunar cycle for the fiducial simulation is to demonstrate the full potential of the instrument, assuming it can survive and operate for that duration.
We also present results for a shorter, more pessimistic, quarter-cycle observation.

For this fiducial run, the instrument turntable was kept at a fixed orientation using the ULSA sky model and the HFSS-derived beams, adding Gaussian random noise with variance given by the radiometer equation.
We therefore assume that static sky is the only present source of signal.
In practice, there will also be time-varying contribution from the Sun and decametric emission from planets.
We discuss how to incorporate such signals into our model in Section \ref{sec:conclusions}.

We note that the current mission baseline includes a 50\% duty cycle for the spectrometer to prevent the battery from discharging below 15\% capacity over the lunar night.
This operational constraint effectively halves the integration time, increasing the radiometric noise by a factor of $\sqrt{2}$ (approximately 41\%).
However, as demonstrated in Section~\ref{sec:results}, our map reconstruction is robust to noise variance increases of up to a factor of 10.
Therefore, this modest increase in noise is not expected to significantly affect the fidelity of the reconstructed maps.

The HFSS simulations include a model of the lunar regolith as a ``layered impedance boundary'' consisting of an eight-layer dielectric slab.
This model mimics the changing properties of the soil with depth, based on a re-analysis of Apollo 17 Surface Electrical Properties (SEP) data~\cite{Grimm2018}.
It assumes a real dielectric constant increasing with depth (from $\sim 3.7$ at the surface to $\sim 6.1$ at 100\,m) and a constant loss tangent of 0.01.
The simulations explicitly account for the reflection from the lunar surface, including ground coupling and interference effects~\cite{Rotermund2026}.

\section{Wiener filter map-making}
\label{sec:wiener}

The observed data vector is assumed to be a linear function of the true sky intensity distribution.
Our model is therefore:
\begin{equation}
        \vd = \mA \vm + \vn,
\end{equation}
where the variables have the following meaning:
\begin{itemize}
        \item $\vd$ is the data vector, a real-valued array constructed by stacking the 4 auto-correlations and the real and imaginary parts of the 6 unique cross-correlations for all observation times.
This ensures that the coupling matrix $\mA$ and the sky map $\vm$ are also real.
        \item $\vm$ is the map vector describing the distribution of intensity on the sky.
This vector could be either a map of sky intensity pixelized in an appropriate spherical pixelization (e.g., \texttt{HEALPix} \cite{2005ApJ...622..759G}), or equivalently a set of spherical harmonics up to some finite $\ell_{\rm max}$ describing the sky intensity in harmonic space.
        \item $\mA$ is the coupling matrix describing how a given distribution of intensity on the sky $\vm$ generates an observed data vector $\vd$.
The sizes of the vectors $\vd$ and $\vm$ are not necessarily (and typically are not) the same;
therefore, the matrix $\mA$ is not, in general, square. The system is well-defined for both under- and over-constrained situations, so $\mA$ can have more rows than columns or vice versa.
        \item $\vn$ is the noise realization, which is assumed to have zero mean ($\left<\vn\right>=0$) and a known covariance matrix $\left<\vn \vn^T\right> =\mN$.
We often assume $\mN$ to be diagonal, but this is not a necessary assumption.
\end{itemize}

This description is quite general and describes any instrument for which non-linearities in the instrument response can be neglected.
In Wiener map-making, we further assume that the map $\vm$ is drawn from a prior probability distribution with zero mean ($\left\langle\vm\right\rangle=0$) and a known signal covariance matrix $\left\langle\vm \vm^T\right\rangle=\mS$.
This prior acts as a regularization, penalizing solutions that deviate significantly from the prior mean, particularly for modes where the data provide weak constraints.
The balance between fitting the data and adhering to this prior is determined by the interplay between the noise covariance $\mN$ and the signal covariance $\mS$.
This framework is mathematically equivalent to Gaussian Process Regression (GPR)~\cite{2006gpml.book.....R}, where the sky map $\vm$ corresponds to the latent function to be inferred, and the signal covariance matrix $\mS$ plays the role of the GPR kernel.

The log posterior for $\vm$ is given by
\begin{equation}
        -2 \log \mathcal{P}(\vm|\vd) = \left(\vd- \mA \vm\right)^T \mN^{-1} (\vd-\mA \vm) + \vm^T\mS^{-1}\vm.
\end{equation}
Maximizing this posterior yields
\begin{equation}
        \hat\vm =  \mW\vd,
        \label{eq:wien}
\end{equation}
with the Wiener filter weighting matrix $\mW$ given by:
\begin{equation}
        \mW = \left(\mA^T \mN^{-1} \mA + \mS^{-1}\right)^{-1}\mA^T \mN^{-1}.
\label{eq:w1}
\end{equation}
Equivalently, the weighting matrix can also be written as (Method 4 of \cite{1997ApJ...480L..87T}):
\begin{equation}
        \mW = \mS \mA^T \left( \mA \mS \mA^T + \mN \right)^{-1}.
\label{eq:w2}
\end{equation}

The weighting matrices in Equations \ref{eq:w1} and \ref{eq:w2} are mathematically equivalent, related by the Woodbury matrix identity.
Both require the inversion of a large square matrix, the dimension of which is determined by the size of the map vector $\vm$ (for Eq.~\ref{eq:w1}) or the data vector $\vd$ (for Eq.~\ref{eq:w2}).
Therefore, given the relative sizes of these two vectors, one formulation might be numerically more advantageous than the other.

By considering the second derivative of the posterior, we can determine the covariance matrix for the best estimate
\begin{equation}
        C_m = \left(\mA^T \mN^{-1} \mA + \mS^{-1}\right)^{-1}.
\label{eq:Cm}
\end{equation}

Similarly to the result for linear least squares, this covariance matrix does not depend on the data vector.

It is important to note that while the true sky intensity is a positive quantity, the Wiener filter is a linear operation and does not enforce this constraint.
This linearity ensures that the estimator remains unbiased; enforcing positivity in noise-dominated regions would rectify the zero-mean noise, introducing a bias.
The resulting maps can therefore have negative values, particularly in regions or on scales where the noise is large and the solution is dominated by the zero-mean prior.
Consequently, the recovery of a positive sky signal serves as a validation that the data is signal-dominated, while regions with statistically significant negative flux would indicate systematic errors.
More advanced non-linear methods would be required to enforce positivity, which is beyond the scope of this paper.

\section{Results}
\label{sec:results}

We applied the Wiener filter map-making algorithm described in Section \ref{sec:wiener} to simulated \LN data.
In this work, we reconstruct the sky map for each frequency channel independently.
Treating frequencies separately ensures the data vector for each reconstruction is sufficiently small to allow for direct numerical implementation of the filter.
The reconstruction was performed in spherical harmonic space, estimating the coefficients $a_{\ell m}$ up to a maximum multipole $\ell_{\rm max} = 47$.
This choice corresponds to an angular resolution limit of approximately $180^\circ / \ell_{\rm max} \approx 3.8^\circ$.

The signal prior covariance matrix $\mS$ was assumed diagonal in harmonic space, with $\mS_{\ell m, \ell' m'} = C_\ell^{\rm prior} \delta_{\ell \ell'} \delta_{m m'}$.
We used the angular power spectrum of the ULSA sky model as the prior power spectrum $C_\ell^{\rm prior}$.
We clarify that this power spectrum is computed directly from the full ULSA sky model without subtracting its mean, resulting in a non-zero power at $\ell=0$, i.e., $C_0 \neq 0$.
This $C_0$ is then used as the variance for the Gaussian prior on the $a_{00}$ mode, which, like all $a_{\ell m}$ priors, is centered at zero.
This approach allows the Wiener filter to robustly reconstruct the sky mean when the variance $C_0$ is sufficiently large.
The specific characteristics of S have a minor impact in the signal-dominated regime.
We employed the formulation in Eq.~\eqref{eq:w1} for the Wiener filter calculation, as the number of map parameters ($N_{\rm modes} \sim \ell_{\rm max}^2$) was smaller than the number of data points ($N_{\rm data}$), making the inversion in map space computationally cheaper.
We present a series of comparisons exploring variations from our fiducial setup.
These variations include the consideration of data from different lunar cycles with varying antenna rotations, different noise levels, and the impact of beam gain fluctuations.

\begin{figure}
        \centering
        \includegraphics[width=\linewidth]{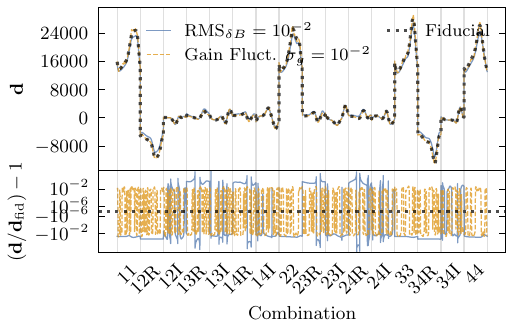}
        \caption{Segment of the simulated visibility data vector $\vd$ for the 25-MHz channel from the fiducial simulation (one lunar cycle, fixed turntable).
The $x$-axis represents a flattened index where visibility data is grouped by antenna pair;
each group forms a segment showing the pair's visibility evolution over all sampled time steps.
Four large sections of the data vector with large positive values correspond to the auto-correlation products.}
        \label{fig:datavector}
\end{figure}

Figure~\ref{fig:datavector} shows a segment of the simulated visibility data vector $\vd$ for the 25-MHz channel.
The $x$-axis of this figure represents the data vector $\vd$ itself, which is constructed as a flattened array of time series measurements.
Specifically, $\vd$ comprises 16 distinct time series, corresponding to the 16 independent real-valued components derived from the antenna correlation products (4 auto-correlations $V_{ii}$, and the real and imaginary parts of the 6 unique cross-correlations $V_{ij}$ with $i<j$).
These 16 time series, each spanning the full observation period, are concatenated sequentially to form $\vd$.
Thus, contiguous segments along the $x$-axis in Figure~\ref{fig:datavector} represent the complete time evolution of individual correlation components, ordered according to the sequence of antenna pairs detailed in the figure caption (e.g., all time samples for $V_{11}$ are followed by all time samples for $\mathrm{Re}(V_{12})$, then $\mathrm{Im}(V_{12})$, and so on).
The data vectors for other frequencies possess qualitatively similar structures.
The observed complexity in $\vd$ arises from the combination of the sky signal and the modulation imposed by the instrument's broad antenna beams.

\subsection{Recovered Sky maps}

\begin{figure*}
        \centering
        \includegraphics[width=\linewidth]{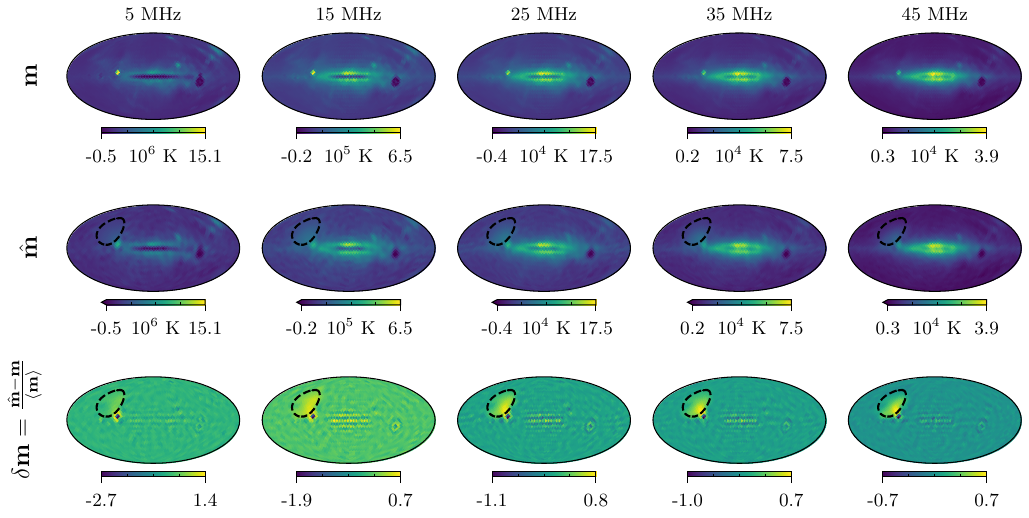}
\caption{Sky map reconstruction results.
Top row: Input ULSA maps smoothed to $\ell_{\rm max}=47$. Middle row: Wiener filter reconstructed maps $\hat{\vm}$ from simulated fiducial data (1 lunar cycle).
Bottom row: difference $\hat{\vm} - \vm$, normalized by the mean temperature of the smoothed truth map.
Maps are shown in Mollweide projection in Galactic coordinates.
The unobserved region within $\sim$24 degrees of the North Celestial Pole, reconstructed purely from the prior, is delimited by a black dashed line in the reconstructed and difference maps.
Prominent dark features correspond to regions of free-free absorption. Residuals (bottom row) are typically at the few-percent level, with larger errors visible near bright features due to limited resolution.
Color scales (units assumed K) are consistent within each frequency column.
Frequencies are 5, 15, 25, 35, and 45 MHz (left to right).
The reconstructed maps may include negative values; these are artifacts from the Wiener filter's regularization in noise-dominated regions.}
        \label{fig:maps_hs}
\end{figure*}

Figure \ref{fig:maps_hs} presents the results of the map reconstruction at several representative frequencies (5, 15, 25, 35 and 45 MHz) in galactic coordinates.
The top row shows the input ULSA ground truth maps, smoothed to the reconstruction resolution ($\ell_{\rm max}=47$).
The middle row displays the Wiener filter reconstructed maps $\hat{\vm}$.
The bottom row shows the absolute difference between the reconstruction and the (smoothed) ground truth, normalized by the mean brightness temperature of the ground truth map at that frequency.

Visually, the Wiener filter successfully recovers the large-scale features of the input sky map across the frequency range.
The bright Galactic Center and Galactic plane are clearly visible, as well as the attenuation coming from free-free absorption~\cite{2022A&A...668A.127P}.
The reconstruction fidelity appears reasonably good, although some differences are noticeable, particularly at lower frequencies where the beam patterns are broadest and potentially more susceptible to modeling errors.
The normalized differences indicate residuals are typically at the few percent level relative to the mean sky temperature.
The reconstruction also demonstrates the filter's behavior where data is absent: a region within approximately 24 degrees of the North Celestial Pole is unobserved by the instrument and is therefore reconstructed purely from the prior.

\subsection{Effective Resolution and Fidelity}

To quantify the reconstruction quality as a function of angular scale, we computed two metrics in harmonic space: the cross-correlation coefficient and the signal-to-noise ratio (SNR).

The cross-correlation coefficient $\rho_\ell$ between the reconstructed map harmonic coefficients $\hat{a}_{\ell m}$ and the true (smoothed) coefficients $a_{\ell m}^{\rm true}$ at multipole $\ell$ is defined as:
\begin{equation}\label{eq:rho_ell}
        \rho_\ell = \frac{\sum_m \hat{a}_{\ell m} (a_{\ell m}^{\rm true})^*}{\sqrt{(\sum_{m} |\hat{a}_{\ell m}|^2) (\sum_{m} |a_{\ell m}^{\rm true}|^2)}}.
\end{equation}

\begin{figure}
        \centering
        \includegraphics[width=\linewidth]{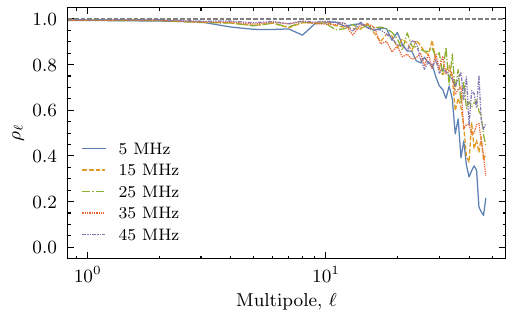}
        \caption{Cross-correlation coefficient $\rho_\ell$ (Eq.~\ref{eq:rho_ell}) between the Wiener filter reconstructed map and the smoothed ground truth map, as a function of spherical harmonic multipole $\ell$.
Results are shown for frequencies from 5 to 50 MHz.
The reconstruction used $\ell_{\rm max}=47$.}
        \label{fig:correlation_hs}
\end{figure}

Figure \ref{fig:correlation_hs} shows the cross-correlation coefficient, $\rho_\ell$, between the Wiener-filtered reconstructed maps and the ground truth ULSA maps as a function of harmonic multipole, $\ell$, for several frequencies.
A value of $\rho_\ell \approx 1$ indicates an excellent reconstruction of the spatial pattern at an angular scale of $\sim 180/\ell$ deg, while $\rho_\ell \approx 0$ signifies no correlation.
The correlation is very high ($\rho_\ell > 0.9$) at the largest scales ($\ell \lesssim 10$) and diminishes towards smaller scales.
This trend is consistent across all frequency bands, with the correlation remaining significant ($\rho_\ell > 0.5$) up to multipoles of $\ell \approx 20-35$.
This suggests that the spatial patterns of structures with angular sizes down to approximately $5$ to $10$ deg are reliably recovered.

While correlation measures pattern similarity, the signal-to-noise ratio (SNR) assesses the amplitude fidelity relative to the residual error.
We define an effective SNR per multipole $\ell$ as the ratio of the power in the reconstructed map to the power in the residual map (reconstruction minus truth),
\begin{equation}\label{eq:snr_ell}
        \mathrm{SNR}_\ell^2 = \frac{\sum_{m} |\hat{a}_{\ell m}|^2}{\sum_{m} |\hat{a}_{\ell m} - a_{\ell m}^{\rm true}|^2}.
\end{equation}

\begin{figure}
        \centering
        \includegraphics[width=\linewidth]{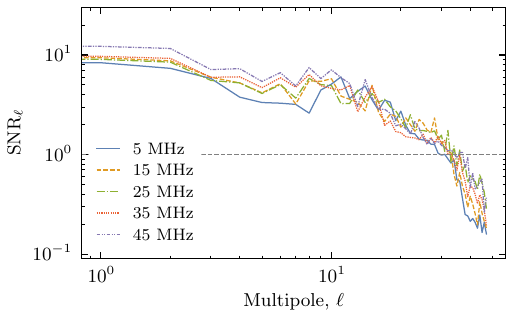}
        \caption{Effective signal-to-noise ratio $\mathrm{SNR}_\ell$ (Eq.~\ref{eq:snr_ell}) of the reconstructed map per multipole $\ell$, plotted on a logarithmic scale.
$\mathrm{SNR}_\ell > 1$ indicates modes where the reconstructed power exceeds the residual error power.
Results are shown for frequencies from 5 to 50 MHz.}
        \label{fig:snr}
\end{figure}

Figure \ref{fig:snr} plots $\mathrm{SNR}_\ell$ versus $\ell$. As expected, the SNR is highest at low $\ell$ (large scales), where the signal is strong and the instrument has good sensitivity.
The SNR decreases with increasing $\ell$ as the signal power drops and the instrument resolution limit is approached.
The SNR remains greater than unity ($\mathrm{SNR}_\ell > 1$) up to $\ell \approx 20$ for all frequency bands, indicating that the reconstructed amplitude is dominated by the true signal rather than reconstruction errors up to these scales.
Beyond $\ell \approx 20$--$35$, the Wiener filter increasingly suppresses the reconstruction towards the prior mean (zero) as the data becomes less constraining.

To test the robustness of our map-making, we consider scenarios with higher noise levels.
We increased the noise covariance, $\mN$, by factors of 10 and 100 compared to the baseline level (\ref{eq:radiometer}).
The sampled noise $\vn$ change accordingly while the underlying sky signal model and the signal prior ($\mS$) remained unchanged in these tests.
The Wiener filter (Eqs. \ref{eq:w1}-\ref{eq:w2}) inherently trusts noisy data less compared to the assumed signal prior ($\mS^{-1}$).
As expected, increasing the noise led to noticeable degradation in the reconstructed maps (not explicitly shown).
With 10 times the baseline noise variance, the maps became smoother and lost finer details.
At 100 times the baseline noise variance, the filter heavily smoothed the map, and only the brightest, largest features like the Galactic plane were clearly recovered.
This confirms that map quality directly depends on the signal-to-noise in the visibility data.
While \LN should operate in a high signal-to-noise regime under nominal conditions, achieving the map fidelity seen in Figure \ref{fig:maps_hs_multicase} relies on maintaining low effective noise, close to the baseline level predicted by the radiometer equation.
We analyze the quantitative impact of noise on different angular scales in the following sections.

Taken together, Figures \ref{fig:correlation_hs} and \ref{fig:snr} indicate that \LN, using the Wiener filter method, should be able to produce reliable sky maps with an effective angular resolution corresponding to $\ell \approx 20$--$35$ (roughly 5--10 degrees) over the 5--50 MHz range, based on one lunar cycle of observation with idealized beams and noise.

\subsection{Gain fluctuations}
\label{sec:gain_fluctuations}

\begin{figure}
        \centering
        \includegraphics[width=\linewidth]{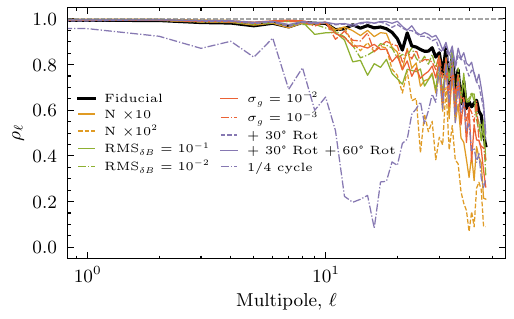}
        \caption{Comparison of reconstruction fidelity ($\rho_\ell$) at 25 MHz for different simulation scenarios.
The plot shows the cross-correlation between the reconstructed map and the ground truth for: the fiducial case (black);
enhanced noise levels (orange); beam model uncertainty (RMS$_{\delta B} \approx 10\%, 1\%$; green); simulated gain fluctuations ($\sigma_g=0.01, 0.001$; red); and extended/reduced observations with turntable rotations (2, 3, and 1/4 lunar cycles; purple).}
        \label{fig:rho_comparison}
\end{figure}

\begin{figure}
        \centering
        \includegraphics[width=\linewidth]{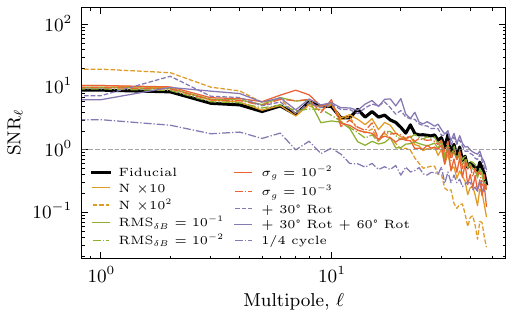}
        \caption{Comparison of reconstruction signal-to-noise ratio ($\mathrm{SNR}_\ell$) at 25 MHz for different simulation scenarios.
The plot shows the SNR per multipole for: the fiducial case (black); enhanced noise levels (orange);
beam model uncertainty (RMS$_{\delta B} \approx 10\%, 1\%$; green); simulated gain fluctuations ($\sigma_g=0.01, 0.001$; red);
and extended/reduced observations with turntable rotations (2, 3, and 1/4 lunar cycles; purple).
Compare with Figure \ref{fig:snr} for the baseline SNR and Figure \ref{fig:rho_comparison} for the cross-correlation comparison.}
        \label{fig:snr_comparison}
\end{figure}

Real instruments experience gain variations over time (e.g., from thermal effects, amplifier drift), which can impact map reconstruction.
We model these as a per-antenna, time-varying Gaussian random process $g_i(t)$ with zero mean, standard deviation $\sigma_g$, and temporal correlation scale $\sigma_t$.
For the scenarios depicted by the red lines in Figures~\ref{fig:rho_comparison} and \ref{fig:snr_comparison}, we set $\sigma_g=0.01$ and $\sigma_t=10$ hours as plausible fiducial values representing slow instrumental drifts due to thermal variations over the course of a lunar day-night cycle.
These fluctuations modify the ideal visibilities $V_{ij}^{\rm ideal}(t)$ to $V_{ij}^{\rm noisy}(t) = (1 + g_i(t))(1 + g_j(t)) V_{ij}^{\rm ideal}(t)$.
The difference, $\epsilon_{ij}(t) = V_{ij}^{\rm noisy}(t) - V_{ij}^{\rm ideal}(t)$, acts as an additional noise source.
This gain-induced noise is not white; it exhibits temporal and inter-visibility correlations.

To optimally account for this structured noise within our Wiener filter framework, we explicitly incorporate its full covariance.
The derivation of this gain noise covariance matrix, $\mN_{\text{gain}} \equiv \mathrm{Cov}(\epsilon_{ij}(t), \epsilon_{kl}(t'))$, is detailed in Appendix~\ref{sec:appendix_gain_cov}, with the final analytical expression given by Eq.~\eqref{eq:cov_epsilon}.
For the map reconstruction results shown in Figures~\ref{fig:rho_comparison} and \ref{fig:snr_comparison} (red lines), the total noise covariance matrix used in the Wiener filter is $\mN = \mN_{\text{radiometer}} + \mN_{\text{gain}}$, where $\mN_{\text{radiometer}}$ is the standard diagonal instrumental noise (Eq.~\eqref{eq:radiometer}).
The coupling matrix $\mA$ remains based on the ideal (mean) antenna gains.

\subsection{Impact of Beam Model Uncertainty}
\label{sec:beam_precision}

Accurate knowledge of the antenna beam patterns $B_{ij}$, which define the coupling matrix $\mA$, is critical for high-fidelity map-making.
Uncertainties in the beam model can introduce significant systematic errors.
These can arise from factors such as imperfect electromagnetic simulations, unmodeled environmental effects (e.g., regolith dielectric properties), mechanical uncertainties like antenna sag, or mispointing of the turntable.
A naive reconstruction that ignores these uncertainties can lead to artifacts and a loss of fidelity.
In fact, irreducible uncertainty in properties of the regolith means that we are unlikely to know the beam response very well, with likely errors in the 1\%-10\% range.
To model plausible beam errors, we cannot simply add white noise to the observed beam, since beam errors will likely be smooth and affect large-scale features.

A more robust approach, analogous to our treatment of gain fluctuations, is to incorporate the impact of beam model uncertainty directly into the noise covariance matrix used by the Wiener filter.
If the true coupling matrix is $\mA_{\text{true}}$ but our reconstruction uses a model $\mA_{\text{model}}$, the data vector can be written as:
\begin{equation}
        \vd = \mA_{\text{true}} \vm + \vn = \mA_{\text{model}} \vm + (\delta \mA \vm) + \vn,
\end{equation}
where $\delta \mA = \mA_{\text{true}} - \mA_{\text{model}}$ is the error in the coupling matrix.
The term $\delta \mA \vm$ represents a systematic error that depends on the true sky signal.
This error can be treated as an additional, structured noise source.

Assuming the true sky map $\vm$ is a Gaussian random field with a prior covariance $\mS$, the covariance of this beam-induced error term, averaged over sky realizations, is given by:
\begin{equation}
        \mN_{\text{beam}} = \langle (\delta \mA \vm) (\delta \mA \vm)^T \rangle = \delta \mA \mS (\delta \mA)^T.
\end{equation}

This covariance can be calculated if a model for the beam uncertainty, $\delta \mA$, is available.
The total effective noise covariance for the Wiener filter then becomes $\mN_{\text{eff}} = \mN_{\text{radiometer}} + \mN_{\text{beam}}$.
By using this augmented noise covariance, the filter appropriately downweights data components that are most susceptible to the modeled beam uncertainties, thereby enhancing the robustness of the reconstruction.

Since departures from ideality for simple antennas such as dipoles are often modelled as an effective change in antenna length, we adopt a similar approach here and simulate uncertainties by introducing a frequency-dependent mismatch.
We generated data using beams from a frequency $f + \Delta f$ but reconstructed the map using the nominal coupling matrix for frequency $f$.
This scenario corresponds to the case where $\delta \mA = \mA(f+\Delta f) - \mA(f)$.
For the 25 MHz channel, we tested the impact of beam model errors using frequency shifts of $\Delta f = 4$ and $0.4$\,MHz, which correspond to beam RMSDs of approximately 10\% and 1\% respectively.
As shown in Figures~\ref{fig:rho_comparison} and \ref{fig:snr_comparison}, such large errors on the beam model do not introduce large discrepancies on the recovered sky map;
they roughly give the same resolution, touching SNR=1 at the same multipole as the fiducial case, but with poorer SNR on intermediate scales, making the maps a bit blurrier.
These results indicate that for realistic levels of uncertainty, incorporating the beam error covariance $\mN_{\text{beam}}$ into the filter is crucial for mitigating systematic artifacts and preserving map fidelity.

\subsection{Impact of Extended Observations and Turntable Rotation}
\label{sec:turntable}

\begin{figure*}
        \centering
        \includegraphics[width=\textwidth]{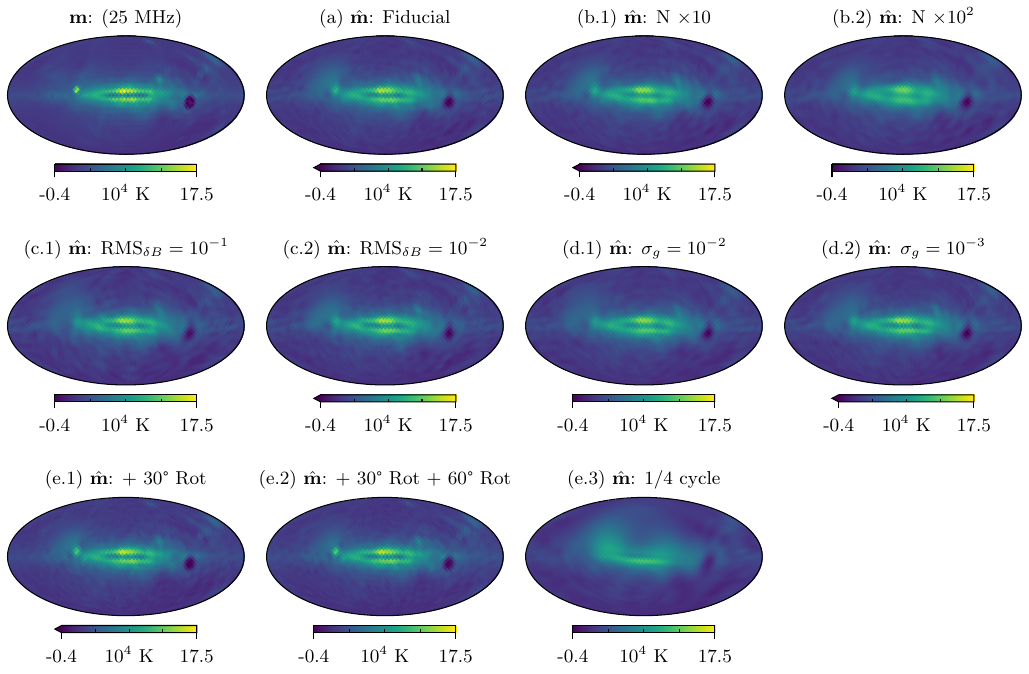}
        \caption{Reconstructed sky maps at 25 MHz (smoothed to $\ell_{\rm max}=47$) for various simulation scenarios.
The grid displays (top-left) the input ULSA ground truth map, followed by Wiener filter reconstructions $\hat{\vm}$ for:
                (a) Fiducial (1 lunar cycle, 0 deg rotation, ideal conditions);
                (b) Noise enhanced: $\times 10$ and $\times 100$;
                (c) Beam uncertainty: RMS$_{\delta B} \approx 10\%$ and $1\%$;
                (d) Gain fluctuations: $\sigma_g = 0.01$ and $\sigma_g = 0.001$;
                (f) Extended observations: 2 cycles with +30 deg rotation, 3 cycles with +30/+60 deg rotations, and a quarter lunar cycle observation.
Details for scenarios (b-f) are relative to the fiducial (a).
                All scenarios are also quantitatively analyzed in Figs.~\ref{fig:rho_comparison} and \ref{fig:snr_comparison}.
Maps are in Mollweide projection (Galactic coordinates) with consistent color scales (K).}
        \label{fig:maps_hs_multicase}
\end{figure*}

The fiducial results presented above were based on a simulation spanning a single lunar cycle with a fixed turntable orientation ($0^\circ$).
To assess the potential benefits of longer observation campaigns and utilizing the instrument's rotational capability, we extended the simulation.
We generated additional data corresponding to two more full lunar cycles, incorporating turntable rotations to $+30$ and $+60$ deg relative to the lander body (one distinct angle per cycle).
This tripled the total volume of simulated data and provided different perspectives on the sky for the antenna system.

We concatenated the data vectors from all three cycles (0, +30, +60 deg rotations) and applied the same Wiener filter map-making procedure, adjusting the noise covariance $\mN$ appropriately for the combined dataset.
The resulting map quality was evaluated by computing the effective signal-to-noise ratio $\mathrm{SNR}_\ell$ as defined in Eq.~\eqref{eq:snr_ell}.

The purple lines in Figures~\ref{fig:rho_comparison} and \ref{fig:snr_comparison} compare the results for extended (2 and 3 cycles) and reduced (1/4 cycle) datasets against the fiducial 1-cycle result (black line).
Two key observations emerge:
i) The ultimate resolution limit, characterized by the multipole $\ell_{\rm cross}$ where $\mathrm{SNR}_\ell$ drops below unity, remains largely unchanged, still occurring around $\ell_{\rm cross} \approx 20-35$.
This result suggests that the finest achievable resolution with this linear method is primarily limited by the intrinsic properties of the instrument (e.g., the broadness of the beams, effective shortest baselines) rather than solely by the integration time, at least within this range of data volume.
ii) However, a significant improvement is observed at intermediate angular scales ($\ell \gtrsim 10$). In this range, the $\mathrm{SNR}_\ell$ for the 3-cycle dataset is a factor of 2-3 higher than for the 1-cycle dataset.

This demonstrates the value of both longer integration times and strategic use of the turntable.
The increased data volume naturally reduces the impact of random noise (variance decreasing roughly as $1/N_{\rm samples}$).
Furthermore, observing the sky with different instrument orientations provides complementary spatial sampling, helping to break degeneracies between different sky modes and improving the overall conditioning of the inverse problem represented by the $\mA^T \mN^{-1} \mA$ term in the Wiener filter solution (Eq.~\ref{eq:w1}).
While estimating the power for fine angular scales remains challenging, the enhanced SNR at intermediate scales implies a more robust and higher-fidelity reconstruction of features with angular sizes corresponding to this range. Further,  we have applied the additional data to our fiducal case. In practice, we will likely have to marginalise over both beam and gain uncertainties and in this case, the additional data might bring less marginal benefits, compensating for the particular subset of modes that are lost due those uncertainties.

Going towards a smaller dataset, we start by noting that intuitively one would assume that a radio transit telescope would need at a minimum a full 360$^\circ$ cycle to recover the full sky.
However, given a telescope that can integrate over $2\pi$ sky area instantaneously with a sufficient number of independent beams, a half-cycle (180$^\circ$) suffices to at least theoretically reconstruct the entire sky.
Over half rotation, each point source on the sky would either raise or settle behind a horizon.
Our fiducial case is therefore already marginally over-constrained. Therefore we study a second case by reconstructing a quarter lunar cycle data.
This would correspond to the pessimistic science achiveable by \LN, assuming the system lasts for half lunar night and then looses sufficient power to survive the night.
Even with a significantly reduced dataset, the largest sky features are discernible and the galaxy is unequivocally detected.
However, as expected, reducing the observation time substantially impacts the reconstruction quality.
Compared to the fiducial full-cycle simulation, the signal-to-noise ratio per multipole ($\mathrm{SNR}_\ell$) is degraded by approximately a factor of five across most scales.
Consequently, the effective resolution---defined as the point where $\mathrm{SNR}_\ell$ crosses unity---is lowered to $\ell \approx 10$.
This result highlights that achieving the full planned observation duration is crucial for maximizing the scientific return in terms of map fidelity and angular resolution.

\section{Conclusions \& Outlook}
\label{sec:conclusions}

We have investigated the feasibility of reconstructing low-frequency radio sky maps from the upcoming \LN instrument on the lunar far side.
Employing a linear map-making algorithm based on the Wiener filter, we demonstrated that \LN's measurement strategy – combining 16 correlation products with sky scanning via lunar rotation over one sidereal period – provides sufficient information to deconvolve the broad antenna beams and recover the sky intensity distribution.

Our simulations, based on the ULSA sky model and HFSS antenna simulations (assuming four-fold symmetry), show that reliable maps can be reconstructed across the 5--50 MHz frequency range.
The effective angular resolution, defined by the multipole range where both the spatial pattern correlation ($\rho_\ell$) is high and the signal-to-noise ratio ($\mathrm{SNR}_\ell$) exceeds unity, corresponds to $\ell \approx 35$.
This value translates to an angular resolution of roughly 5 degrees.
The Wiener filter successfully recovers the main features of the Galactic synchrotron emission, providing a valuable dataset for Galactic science and potentially for characterizing foregrounds for cosmological studies.

We performed preliminary assessments of the impact of instrumental systematics.
While uncalibrated time-varying gain fluctuations and imperfect beam knowledge can degrade reconstruction fidelity, the adopted Wiener filter framework demonstrates resilience.
Specifically, by incorporating the full covariance of gain-induced noise (as detailed in Appendix~\ref{sec:appendix_gain_cov}), our analysis suggests that the target map quality is achievable provided residual gain fluctuations are controlled or characterized to the order of 1--10\%, and the \emph{in situ} beam patterns are determined with an accuracy of approximately 10\%.

Furthermore, we investigated the benefits of extending the observation duration and utilizing the instrument's turntable capability.
Simulating three lunar cycles with distinct turntable orientations (0, +30, +60 deg) demonstrated a significant improvement in the signal-to-noise ratio for intermediate angular scales ($\ell \gtrsim 10$) compared to the single-cycle, fixed-orientation baseline.
While the finest achievable angular resolution ($\ell \approx 20-35$, where SNR drops below unity) remained largely unchanged, the enhanced fidelity at intermediate scales highlights the value of accumulating more data and leveraging different viewing geometries.

There are natural extensions of this work within the formalism of Wiener filtering:
\begin{enumerate}
        \item There will be point sources on the sky, such as Fornax A, which will be unresolved by \LN.
Since we likely know the position of those sources, a better linear model would contain a small number of known point sources, whose fluxes are fitted parameters in addition to the diffuse emission modelled as a map.
First work in this direction has already started~\cite{Fahs2026}.

        \item In a similar vein, it should also be possible to add sources with more complex polarization structures and those that are moving on the celestial sphere -- prime examples being Jupiter, with its circularly polarized decametric emission, and the Sun.
        \item An additional extension is to model multiple frequencies. We know that neighboring frequencies will be morphologically strongly correlated.
The same method can be applied by reconstructing all frequencies at once by encoding this similarity into the prior covariance matrix $\mS$.
The main issue with this approach is that the size of the data vectors would increase beyond what can be dealt with using brute-force methods.
This would require a significantly more sophisticated approach than the one taken in this paper, potentially using techniques like consensus optimization to manage the computational scale (e.g., \cite{2016arXiv160509219Y}).
        \item A fourth avenue is to move beyond linear methods by jointly solving for the sky map and parameterized instrumental effects.
For small perturbations, such as minor gain drifts ($\delta g$) or beam errors ($\delta \mA$), an iterative approach could be employed to refine both the map and the instrument model.
This highlights the primary limitation of a purely linear framework: the assumption that the coupling matrix $\mA$ is sufficiently well-known. Non-linear methods allow for a joint solution that can account for complex systematic effects that are difficult to model as simple additive noise.
\end{enumerate}

Beyond linear methods, there is significant interest in expanding this work into non-linear fitting by introducing a general parameterized model of both the instrument and the sky and using direct forward modelling to constrain it.
Such a model can have significantly fewer parameters and can capture non-linear uncertainties in the beam shape that can be self-calibrated, perhaps with the help of a weakly informative prior.
Such models can, however, be biased towards the sky as we expect it, either through informative priors or priors implicit in the model itself.
Therefore, Wiener deconvolution should be viewed as a statistically efficient, regularized (and thus biased in low-SNR modes) method to compress the observed data vector into a visual representation.

In conclusion, our analysis indicates that \LN has the potential to deliver very valuable low-resolution maps of the low-frequency radio sky from the unique vantage point of the lunar far side, provided that instrumental characteristics are reasonably well understood.

\begin{acknowledgments}
This research was supported by the U.S. Department of Energy, Office of Science under the LuSEE-Night Science program and NASA under contract 80GSFC21C0011.
L.V.E.K. acknowledges support from the European Research Council (ERC) under the European Union's Horizon 2020 research and innovation programme, grant agreement 884760 (CoDEX).
P. Zarka acknowledges funding from the European Research Council (ERC) under the European Union's Horizon 2020 research and innovation programme (grant agreement No 101020459 - Exoradio).
D.R acknowledges funding from NASA APRA grant award 80NSSC23K0013.
\end{acknowledgments}

\bibliography{main.bib}

@article{2025arXiv250309842H,
  author        = {{Hibbard}, Joshua J. and {Burns}, Jack O. and {MacDowall}, Robert and {Gopalswamy}, Natchimuthuk and {Boardsen}, Scott A. and {Farrell}, William and {Bradley}, Damon and {Schulszas}, Thomas M. and {Dorigo Jones}, Johnny and {Rapetti}, David and {Turner}, Jake D.},
  title         = {{Results from NASA's First Radio Telescope on the Moon: Terrestrial Technosignatures and the Low-Frequency Galactic Background Observed by ROLSES-1 Onboard the Odysseus Lander}},
  journal       = {arXiv e-prints},
  year          = 2025,
  month         = mar,
  eid           = {arXiv:2503.09842},
  pages         = {arXiv:2503.09842},
  doi           = {10.48550/arXiv.2503.09842},
  archiveprefix = {arXiv},
  eprint        = {2503.09842},
  primaryclass  = {astro-ph.IM},
  adsurl        = {https://ui.adsabs.harvard.edu/abs/2025arXiv250309842H}
}

@article{2022A&A...668A.127P,
  author   = {{Page}, B. and {Bassett}, N. and {Lecacheux}, A. and {Pulupa}, M. and {Rapetti}, D. and {Bale}, S.~D.},
  title    = {{The l = 2 spherical harmonic expansion coefficients of the sky brightness distribution between 0.5 and 7 MHz}},
  journal  = {\aap},
  year     = 2022,
  month    = dec,
  volume   = {668},
  eid      = {A127},
  pages    = {A127},
  doi      = {10.1051/0004-6361/202244621},
  adsurl   = {https://ui.adsabs.harvard.edu/abs/2022A&A...668A.127P}
}

@article{2016arXiv160509219Y,
  author        = {{Yatawatta}, Sarod},
  title         = {{Fine tuning consensus optimization for distributed radio interferometric calibration}},
  journal       = {arXiv e-prints},
  year          = 2016,
  month         = may,
  eid           = {arXiv:1605.09219},
  pages         = {arXiv:1605.09219},
  doi           = {10.48550/arXiv.1605.09219},
  archiveprefix = {arXiv},
  eprint        = {1605.09219},
  primaryclass  = {astro-ph.IM},
  adsurl        = {https://ui.adsabs.harvard.edu/abs/2016arXiv160509219Y}
}

@book{2006gpml.book.....R,
  author  = {{Rasmussen}, Carl Edward and {Williams}, Christopher K.~I.},
  title   = {{Gaussian Processes for Machine Learning}},
  year    = 2006,
  adsurl  = {https://ui.adsabs.harvard.edu/abs/2006gpml.book.....R}
}

@article{2025arXiv250403418B,
  author        = {{Brinkerink}, C.~D. and {Arts}, M.~J. and {Bentum}, M.~J. and {Boonstra}, A.~J. and {Cecconi}, B. and {Fialkov}, A. and {Garcia Guti{\'e}rrez}, J. and {Ghosh}, S. and {Grenouilleau}, J. and {Gurvits}, L.~I. and {Klein-Wolt}, M. and {Koopmans}, L.~V.~E. and {Lazendic-Galloway}, J. and {Paragi}, Z. and {Prinsloo}, D. and {Rajan}, R.~T. and {Rouill{\'e}}, E. and {Ruiter}, M. and {Tauber}, J.~A. and {Vedantham}, H.~K. and {Vecchio}, A. and {Vertegaal}, C.~J.~C. and {Zandboer}, J.~C.~F. and {Zucca}, P.},
  title         = {{The Dark Ages Explorer (DEX): a filled-aperture ultra-long wavelength radio interferometer on the lunar far side}},
  journal       = {arXiv e-prints},
  year          = 2025,
  month         = apr,
  eid           = {arXiv:2504.03418},
  pages         = {arXiv:2504.03418},
  doi           = {10.48550/arXiv.2504.03418},
  archiveprefix = {arXiv},
  eprint        = {2504.03418},
  primaryclass  = {astro-ph.IM},
  adsurl        = {https://ui.adsabs.harvard.edu/abs/2025arXiv250403418B}
}

@unpublished{Li2026,
  author = {{Li}, Zack and others},
  year   = {2026},
  note   = {in preparation}
}

@unpublished{Fahs2026,
  author = {{Fahs}, Adam and others},
  year   = {2026},
  note   = {in preparation}
}

@article{1998ApJ...503..492S,
  author        = {{Seljak}, Uro{\v{s}}},
  title         = {{Cosmography and Power Spectrum Estimation: A Unified Approach}},
  journal       = {\apj},
  year          = 1998,
  month         = aug,
  volume        = {503},
  number        = {2},
  pages         = {492-501},
  doi           = {10.1086/306019},
  archiveprefix = {arXiv},
  eprint        = {astro-ph/9710269},
  primaryclass  = {astro-ph},
  adsurl        = {https://ui.adsabs.harvard.edu/abs/1998ApJ...503..492S}
}

@inproceedings{lcrt,
  author    = {Bandyopadhyay, Saptarshi and Mcgarey, Patrick and Goel, Ashish and Rafizadeh, Ramin and Delapierre, Melanie and Arya, Manan and Lazio, Joseph and Goldsmith, Paul and Chahat, Nacer and Stoica, Adrian and Quadrelli, Marco and Nesnas, Issa and Jenks, Kenneth and Hallinan, Gregg},
  booktitle = {2021 IEEE Aerospace Conference (50100)},
  title     = {Conceptual Design of the Lunar Crater Radio Telescope (LCRT) on the Far Side of the Moon},
  year      = {2021},
  volume    = {},
  number    = {},
  pages     = {1-25},
  doi       = {10.1109/AERO50100.2021.9438165}
}

@inproceedings{2019EPSC...13..529S,
  author    = {{Su}, Yan and {Li}, Chunlai and {Li}, Junduo and {Liu}, Bin and {Yan}, Wei and {Zhu}, Xinying and {Zhang}, Tao},
  title     = {{In-orbit Performance of Chang'e-4 Low Frequency Radio Spectrometer}},
  booktitle = {EPSC-DPS Joint Meeting 2019},
  year      = 2019,
  volume    = {2019},
  month     = sep,
  eid       = {EPSC-DPS2019-529},
  pages     = {EPSC-DPS2019-529},
  adsurl    = {https://ui.adsabs.harvard.edu/abs/2019EPSC...13..529S}
}

@article{2021PSJ.....2...44B,
  author        = {{Burns}, Jack O. and {MacDowall}, Robert and {Bale}, Stuart and {Hallinan}, Gregg and {Bassett}, Neil and {Hegedus}, Alex},
  title         = {{Low Radio Frequency Observations from the Moon Enabled by NASA Landed Payload Missions}},
  journal       = {\psj},
  year          = 2021,
  month         = apr,
  volume        = {2},
  number        = {2},
  eid           = {44},
  pages         = {44},
  doi           = {10.3847/PSJ/abdfc3},
  archiveprefix = {arXiv},
  eprint        = {2102.02331},
  primaryclass  = {astro-ph.IM},
  adsurl        = {https://ui.adsabs.harvard.edu/abs/2021PSJ.....2...44B}
}

@book{2016era..book.....C,
  author  = {{Condon}, James J. and {Ransom}, Scott M.},
  title   = {{Essential Radio Astronomy}},
  year    = 2016,
  adsurl  = {https://ui.adsabs.harvard.edu/abs/2016era..book.....C}
}

@article{1997ApJ...480L..87T,
  author        = {{Tegmark}, Max},
  title         = {{How to Make Maps from Cosmic Microwave Background Data without Losing Information}},
  journal       = {\apjl},
  year          = 1997,
  month         = may,
  volume        = {480},
  number        = {2},
  pages         = {L87-L90},
  doi           = {10.1086/310631},
  archiveprefix = {arXiv},
  eprint        = {astro-ph/9611130},
  primaryclass  = {astro-ph},
  adsurl        = {https://ui.adsabs.harvard.edu/abs/1997ApJ...480L..87T}
}

@article{2023arXiv230110345B,
  author        = {{Bale}, Stuart D. and {Bassett}, Neil and {Burns}, Jack O. and {Dorigo Jones}, Johnny and {Goetz}, Keith and {Hellum-Bye}, Christian and {Hermann}, Sven and {Hibbard}, Joshua and {Maksimovic}, Milan and {McLean}, Ryan and {Monsalve}, Raul and {O'Connor}, Paul and {Parsons}, Aaron and {Pulupa}, Marc and {Pund}, Rugved and {Rapetti}, David and {Rotermund}, Kaja M. and {Saliwanchik}, Ben and {Slosar}, Anze and {Sundkvist}, David and {Suzuki}, Aritoki},
  title         = {{LuSEE 'Night': The Lunar Surface Electromagnetics Experiment}},
  journal       = {arXiv e-prints},
  year          = 2023,
  month         = jan,
  eid           = {arXiv:2301.10345},
  pages         = {arXiv:2301.10345},
  doi           = {10.48550/arXiv.2301.10345},
  archiveprefix = {arXiv},
  eprint        = {2301.10345},
  primaryclass  = {astro-ph.IM},
  adsurl        = {https://ui.adsabs.harvard.edu/abs/2023arXiv230110345B}
}

@article{2021ApJ...914..128C,
  author        = {{Cong}, Yanping and {Yue}, Bin and {Xu}, Yidong and {Huang}, Qizhi and {Zuo}, Shifan and {Chen}, Xuelei},
  title         = {{An Ultralong-wavelength Sky Model with Absorption Effect}},
  journal       = {\apj},
  year          = 2021,
  month         = jun,
  volume        = {914},
  number        = {2},
  eid           = {128},
  pages         = {128},
  doi           = {10.3847/1538-4357/abf55c},
  archiveprefix = {arXiv},
  eprint        = {2104.03170},
  primaryclass  = {astro-ph.GA},
  adsurl        = {https://ui.adsabs.harvard.edu/abs/2021ApJ...914..128C}
}

@article{2021arXiv210308623B,
  author        = {{Burns}, Jack and {Hallinan}, Gregg and {Chang}, Tzu-Ching and {Anderson}, Marin and {Bowman}, Judd and {Bradley}, Richard and {Furlanetto}, Steven and {Hegedus}, Alex and {Kasper}, Justin and {Kocz}, Jonathan and {Lazio}, Joseph and {Lux}, Jim and {MacDowall}, Robert and {Mirocha}, Jordan and {Nesnas}, Issa and {Pober}, Jonathan and {Polidan}, Ronald and {Rapetti}, David and {Romero-Wolf}, Andres and {Slosar}, An{\v{z}}e and {Stebbins}, Albert and {Teitelbaum}, Lawrence and {White}, Martin},
  title         = {{A Lunar Farside Low Radio Frequency Array for Dark Ages 21-cm Cosmology}},
  journal       = {arXiv e-prints},
  year          = 2021,
  month         = mar,
  eid           = {arXiv:2103.08623},
  pages         = {arXiv:2103.08623},
  doi           = {10.48550/arXiv.2103.08623},
  archiveprefix = {arXiv},
  eprint        = {2103.08623},
  primaryclass  = {astro-ph.IM},
  adsurl        = {https://ui.adsabs.harvard.edu/abs/2021arXiv210308623B}
}

@article{2005ApJ...622..759G,
  author        = {{G{\'o}rski}, K.~M. and {Hivon}, E. and {Banday}, A.~J. and {Wandelt}, B.~D. and {Hansen}, F.~K. and {Reinecke}, M. and {Bartelmann}, M.},
  title         = {{HEALPix: A Framework for High-Resolution Discretization and Fast Analysis of Data Distributed on the Sphere}},
  journal       = {\apj},
  year          = 2005,
  month         = apr,
  volume        = {622},
  number        = {2},
  pages         = {759-771},
  doi           = {10.1086/427976},
  archiveprefix = {arXiv},
  eprint        = {astro-ph/0409513},
  primaryclass  = {astro-ph},
  adsurl        = {https://ui.adsabs.harvard.edu/abs/2005ApJ...622..759G}
}

@article{2021RSPTA.37990566C,
  author        = {{Chen}, Xuelei and {Yan}, Jingye and {Deng}, Li and {Wu}, Fengquan and {Wu}, Lin and {Xu}, Yidong and {Zhou}, Li},
  title         = {{Discovering the sky at the longest wavelengths with a lunar orbit array}},
  journal       = {Philosophical Transactions of the Royal Society of London Series A},
  year          = 2021,
  month         = jan,
  volume        = {379},
  number        = {2188},
  eid           = {20190566},
  pages         = {20190566},
  doi           = {10.1098/rsta.2019.0566},
  archiveprefix = {arXiv},
  eprint        = {2007.15794},
  primaryclass  = {astro-ph.IM},
  adsurl        = {https://ui.adsabs.harvard.edu/abs/2021RSPTA.37990566C}
}

@unpublished{Rotermund2026,
  author = {{Rotermund}, Kaja M. and others},
  year   = {2026},
  note   = {in preparation}
}

@article{Grimm2018,
  author   = {{Grimm}, Robert E.},
  title    = {{New analysis of the Apollo 17 surface electrical properties experiment}},
  journal  = {\icarus},
  year     = 2018,
  month    = nov,
  volume   = {314},
  pages    = {389-399},
  doi      = {10.1016/j.icarus.2018.06.007},
  adsurl   = {https://ui.adsabs.harvard.edu/abs/2018Icar..314..389G}
}

\appendix

\section{Gain fluctuations noise covariance}
\label{sec:appendix_gain_cov}

Noise from gain fluctuations is modeled by considering the gain $g_i(t)$ for each antenna $i$ as a time-varying, zero-mean Gaussian random process.
These fluctuations $g_i(t)$ are characterized by:
\begin{itemize}
        \item Independence between antennas: $\langle g_i(t)g_j(t') \rangle = 0$ for $i \neq j$.
        \item Temporal auto-covariance for a single antenna $i$: $\langle g_i(t)g_i(t') \rangle = C_g(t,t')$.
This function models the temporal correlation, with amplitude given by the gain fluctuations variance, $C_g(t,t) = \sigma_g^2$.
We model it by applying a Gaussian smoothing filter with correlation time $\sigma_t$ to white noise (Section~\ref{sec:gain_fluctuations}), resulting in:
              \begin{equation}\label{eq:Cg_definition}
                      C_g(t,t') = \sigma_g^2 \exp\left(-\frac{(t-t')^2}{2\sigma_t^2}\right).
              \end{equation}
\end{itemize}

The gain-corrupted data for the antenna pair $(i,j)$ is then given by $d^{\mathrm{noisy}}_{ij}(t) = (1 + g_i(t))(1 + g_j(t)) d_{ij}(t)$.
So the noise contribution due to these fluctuations is
\begin{equation}
        \epsilon_{ij}(t) = d^{\mathrm{noisy}}_{ij}(t) - d_{ij}(t) = X_{ij}(t) d_{ij}(t),
\end{equation}
where
\begin{equation}\label{eq:epsilon_definition}
        X_{ij}(t) = g_i(t) + g_j(t) + g_i(t) g_j(t).
\end{equation}

The covariance of this noise term can be then computed as
\begin{align}\label{eq:cov_epsilon_general_definition}
        \mathrm{Cov}(\epsilon_{ij}(t), \epsilon_{kl}(t'))  & = d_{ij}(t)d_{kl}(t') \times \notag                                    \\
        \left[ \langle X_{ij}(t)X_{kl}(t') \rangle \right.
        & - \left. \langle X_{ij}(t) \rangle \langle X_{kl}(t') \rangle \right],
\end{align}
where the different terms can be computed from the statistical properties of the gain fluctuations, $g_i(t)$.
The mean of $X_{ij}(t)$ is
$\langle X_{ij}(t) \rangle = \langle g_i(t) \rangle + \langle g_j(t) \rangle + \langle g_i(t)g_j(t) \rangle = \sigma_g^2 \delta_{ij}$.
Thus,
\begin{equation}\label{eq:product_of_means}
        \langle X_{ij}(t) \rangle \langle X_{kl}(t') \rangle = \sigma_g^4 \delta_{ij}\delta_{kl}.
\end{equation}

To evaluate $\langle X_{ij}(t)X_{kl}(t') \rangle$, we expand the product using Eq.~\eqref{eq:epsilon_definition}.
Terms with an odd number of $g$ factors vanish since $\langle g_k(t) \rangle=0$.
The non-zero contributions are then
\begin{align*}
        \langle X_{ij}(t)X_{kl}(t') \rangle
         & = \langle g_i(t)g_k(t') \rangle + \langle g_i(t)g_l(t') \rangle \notag       \\
         & \quad + \langle g_j(t)g_k(t') \rangle + \langle g_j(t)g_l(t') \rangle \notag \\
         & \quad + \langle g_i(t)g_j(t)g_k(t')g_l(t') \rangle.
\end{align*}
The sum of expectations for the terms with two $g$ factors is
\begin{equation}\label{eq:two_g_factor_term}
        C_g(t,t')(\delta_{ik} + \delta_{il} + \delta_{jk} + \delta_{jl}).
\end{equation}
For the four-$g$ factor term, Wick's (Isserlis') theorem yields:
\begin{align}
         & \langle g_i(t)g_j(t)g_k(t')g_l(t') \rangle \notag             \\
         & \quad = \langle g_i(t)g_j(t) \rangle \langle g_k(t')g_l(t') \rangle + \langle g_i(t)g_k(t') \rangle \langle g_j(t)g_l(t') \rangle \notag    \\
         & \qquad + \langle g_i(t)g_l(t') \rangle \langle g_j(t)g_k(t') \rangle \notag
                         \\
        % &\quad = (C_g(t,t)\delta_{ij})(C_g(t',t')\delta_{kl}) \notag \\
        % &\qquad + (C_g(t,t')\delta_{ik})(C_g(t,t')\delta_{jl}) \notag \\
        % &\qquad + (C_g(t,t')\delta_{il})(C_g(t,t')\delta_{jk}) \notag \\
         & \quad = \sigma_g^4\delta_{ij}\delta_{kl} + C_g(t,t')^2 (\delta_{ik}\delta_{jl} + \delta_{il}\delta_{jk}).
\label{eq:four_g_factor_term_final}
\end{align}
Combining Eqs.~\eqref{eq:two_g_factor_term} and~\eqref{eq:four_g_factor_term_final}:
\begin{align}\label{eq:mean_X_X_prime_final_expression}
        \langle X_{ij}(t)X_{kl}(t') \rangle
         & = C_g(t,t')(\delta_{ik} + \delta_{il} + \delta_{jk} + \delta_{jl}) \notag \\
         & \quad + \sigma_g^4 \delta_{ij}\delta_{kl} \notag                            \\
         & \quad + C_g(t,t')^2 (\delta_{ik}\delta_{jl} + \delta_{il}\delta_{jk}).
\end{align}
Substituting Eqs.~\eqref{eq:product_of_means} and~\eqref{eq:mean_X_X_prime_final_expression} into Eq.~\eqref{eq:cov_epsilon_general_definition}, the $\sigma_g^4\delta_{ij}\delta_{kl}$ term cancels, yielding the noise covariance:
\begin{align}\label{eq:cov_epsilon}
         & \mathrm{Cov}(\epsilon_{ij}(t), \epsilon_{kl}(t')) = d_{ij}(t)d_{kl}(t') C_g(t,t') \times                                                           \\
         & \quad \left[ (\delta_{ik} + \delta_{il} + \delta_{jk} + \delta_{jl}) + C_g(t,t') (\delta_{ik}\delta_{jl} + \delta_{il}\delta_{jk}) \right].
 \notag
\end{align}

The covariance structure presented in Eq.~\eqref{eq:cov_epsilon} consists of terms that are linear and quadratic with respect to $C_g(t,t')$.
Given that $C_g(t,t')$ is proportional to $\sigma_g^2$ (Eq.~\eqref{eq:Cg_definition}), the variance of the gain fluctuations $g_i(t)$, the linear term is expected to dominate when these fluctuations are small (i.e., $\sigma_g \ll 1$).
Nevertheless, our analysis utilizes the complete expression.

\end{document}